\documentclass[fleqn,twoside]{article}
\usepackage{espcrc2}
\usepackage{graphicx,color}

\usepackage[hypertex]{hyperref}

\newcommand{\BulletItem}[1]{\noindent$\bullet$\hspace{0.4em}\textbf{#1}}

\newcommand{\PreprintNumber}{Fermilab CONF-01/310-T}
\newcommand{\BtoDstar}{\ensuremath{\bar{B} \to D^* \ell \bar{\nu}}}
\newcommand{\Vcb}{\ensuremath{|V_{cb}|}}
\newcommand{\hAone}{\ensuremath{h_{A_1}(1)}}
\newcommand{\etal}{\textit{et al.}}
\newcommand{\Aerr}[2]{\ensuremath{\vphantom{0}^{+#1}_{-#2}}}

\newcommand{\JournalRef}[4]{{#1}, \textbf{#2} ({#4}) {#3}}

\newcommand{\PRD}{Phys.\ Rev.\ D}

\newcommand{\NucPhys}{Nucl.\ Phys.}

\newcommand{\PLB}{Phys.\ Lett.\ B}

\newcommand{\arXiv}[2]{\href{http://xxx.lanl.gov/abs/#1/#2}{#1/#2}}

\newcommand{\FullStop}{.}

\newcommand{\GeV}{\textrm{GeV}}

\newcommand{\braOket}[3]{%
\ensuremath{\left\langle{\,#1\,}\mid{\,#2\,}\mid{\,#3\,}\right\rangle}}

\newcommand{\TheCombinedResult}{$\hAone=0.913\Aerr{24}{17}\Aerr{17}{30}$}

\newcommand{\ErrBudget}{%
\begin{table}
	\centering
	\caption[tab:budget]{Budget of statistical and systematic
		uncertainties for $\hAone$ and $1-\hAone$.
		The row labeled ``total systematic'' does not include uncertainty
		from fitting, which is lumped with the statistical error.
		The statistical error is that after chiral extrapolation.}
	\label{tab:budget}
	\begin{tabular}{lcc} \hline
		uncertainty &\hAone &$1-\hAone$  \\
		            &       &\% \\
		\hline
	stats and fits			 &$+0.0238-\!0.0173$	&$+27-\!20$ \\
	adj. $m_c$ \& $m_b$        	 &$+0.0066-\!0.0068$	&$+~8-\!~8$ \\
	$\alpha_s^2$			 &$\pm0.0082$		&$\pm~9$  \\
	$\alpha_s(\bar{\Lambda}/2m_Q)^2$ &$\pm0.0114$		&$\pm13$  \\
	$(\bar{\Lambda})^3/(2m_Q)^3$     &$\pm0.0017$		&$\pm~2$  \\
	$a$ dependence                   &$+0.0032-\!0.0141$	&$+~4-\!16$ \\
	chiral extrap.                   &$+0.0000-\!0.0163$	&$+~0-\!19$ \\
	quenching                        &$+0.0061-\!0.0143$	&$+~7-\!16$ \\
	\hline
	total syst.			 &$+0.0171-\!0.0302$	&$+20-\!35$ \\
	stat $\oplus$ syst		 &$+0.0293-\!0.0349$	&$+34-\!40$ \\
	\end{tabular}
\end{table}%
}

\newcommand{\FigSpacDep}{%
\begin{figure}[htb]
\label{fig:LatSpac}%
\includegraphics[clip=true,height=\columnwidth,angle=-90]{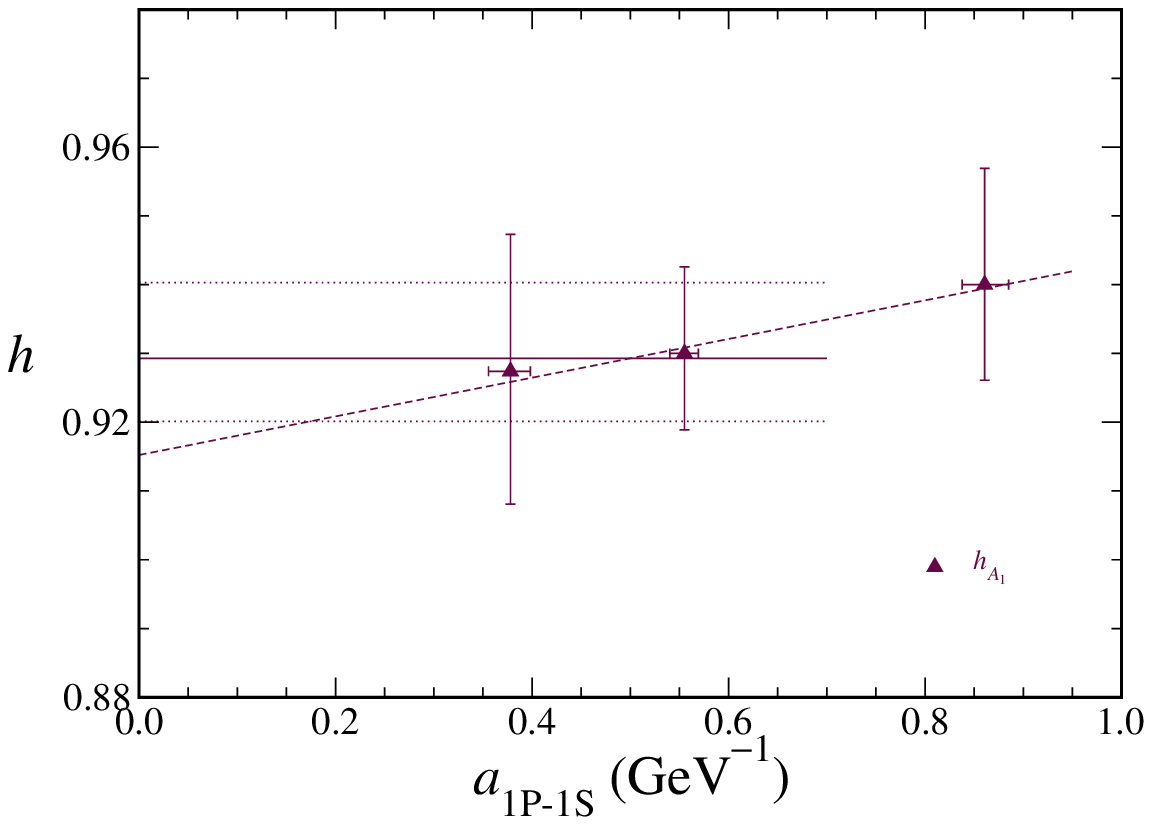}%
\vspace{-5.5ex}
\caption{Lattice spacing dependence for \protect\hAone.}
\end{figure}}

\newcommand{\FigPiDep}{%
\begin{figure}[htb]
\label{fig:SpectMass}%
\includegraphics[clip=true,width=\columnwidth]{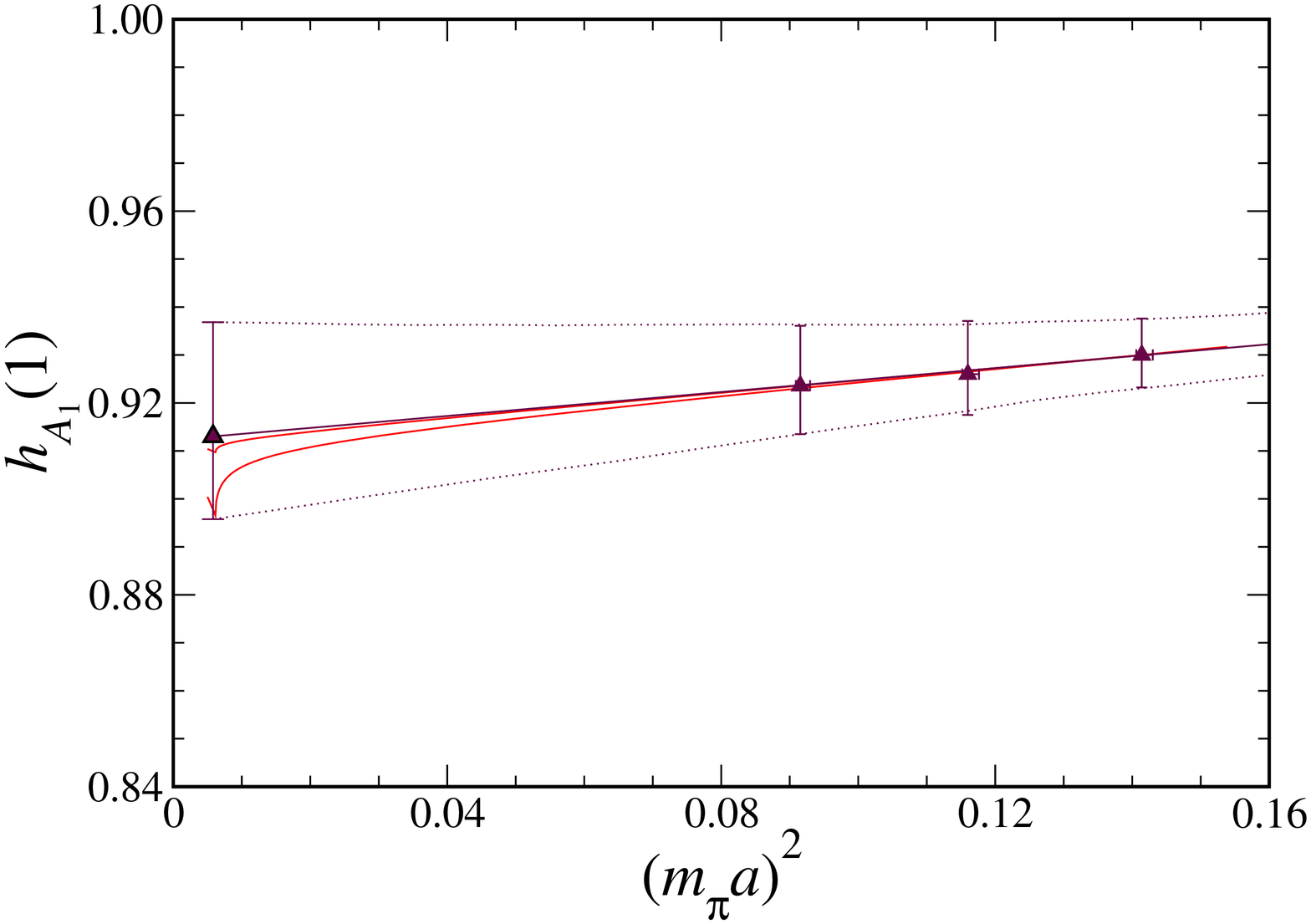}%
\vspace{-5.5ex}
\caption{Spectator quark mass dependence at $\beta=5.9$.}
\end{figure}}

\def\myauthor[#1]#2{{#2}$^{\textrm{#1}}$}
\def\myaddress[#1]#2{$^\textrm{#1}${\textit{#2}}}

\title{\mbox{%
The $\BtoDstar$ form factor at zero recoil
and the determination of $\Vcb$}}

\author{%
\myauthor[a]{J.N.\ Simone}\thanks{Presenter; \texttt{simone@fnal.gov}},
\myauthor[b]{S.\ Hashimoto},
\myauthor[a]{A.S.\ Kronfeld},
\myauthor[a]{P.B.\ Mackenzie}
and \myauthor[c]{S.M.\ Ryan}
\\ [2.8ex]
\myaddress[a]{Fermi National Accelerator Laboratory,
Batavia, IL 60510, USA} \\ [1.2ex]
\myaddress[b]{High Energy Accelerator Research Organization (KEK),
Tsukuba, 305-0801, Japan} \\  [1.2ex]
\myaddress[c]{Department of Applied Mathematics, Trinity College, Dublin 2, Ireland}
}

\begin{document}

\begin{abstract}
We summarize our lattice  QCD study of  the form factor at zero recoil
in the decay \BtoDstar.  After careful consideration of all sources of
systematic uncertainty, we find,  \TheCombinedResult, where the  first
uncertainty is  from statistics and fitting  while the second combined
uncertainty is from all other systematic effects.
\end{abstract}

\maketitle

\section{INTRODUCTION}

A precise   value  for  CKM  matrix  element $\Vcb$  is  an  important
ingredient  in the  study  of CP  violation  and  the determination of
Wolfenstein parameters  $(\bar{\rho},\bar{\eta})$ via   the  unitarity
triangle.

Experimental   studies of  the  $\BtoDstar$  decay rate determine  the
combination  $\Vcb\,\hAone$\cite{CLEO,LEP-Vcb-WG,Belle}.  The hadronic
form factor $\hAone$  at zero recoil  must be computed by  theoretical
means in order to extract $\Vcb$.

Heavy  quark symmetry   imposes powerful constraints   on $\hAone$: it
requires  $\hAone\to1$ in the  infinite mass  limit\cite{Isgur} and it
determines the  structure  of $1/m_Q^p$ corrections\cite{Luke}.  These
power corrections correspond to long distance matrix elements in Heavy
Quark Effective Theory  (HQET).   Previous determinations of  $\hAone$
have employed  sum  rules\cite{Shifman}  or  have appealed    to quark
models\cite{Neubert}.

We present a  determination of $\hAone$ from  lattice QCD using a  new
double  ratio method   designed  so   that  the  bulk of    correlated
statistical  and systematic  errors   cancel\cite{BtoDstar}.  In fact,
with  this method nearly  all errors --   including quenching -- scale
with $1-h_{A_1}$ rather than with $h_{A_1}$.

Our  result is computed   in  the quenched approximation.  The  method
lends itself to    a  completely model  independent  determination  of
$\hAone$ once unquenched  gauge configurations become  available.  The
full details of our analysis are found in Reference~\cite{BtoDstar}.

\section{THE METHOD}

In the heavy quark expansion\cite{BtoDstar},
\begin{eqnarray}
\hAone=\eta_A\left[
1
-\frac{\ell_V}{(2m_c)^2} -\frac{\ell_V^{(3)}}{(2m_c)^3}
-\frac{\ell_P}{(2m_b)^2}\right.
&& \nonumber \\
 -\frac{\ell_P^{(3)}}{(2m_b)^3}
+\frac{2\ell_A}{2m_c2m_b}
+\frac{\ell_A^{(3)}}{2m_c2m_b}\left(\frac{1}{2m_c}+\frac{1}{2m_b}\right) && \nonumber \\
\left.
+\frac{\ell_-^{(3)}}{2m_c2m_b}\left(\frac{1}{2m_c}-\frac{1}{2m_b}\right)
+\cdots\; \right]\FullStop\qquad(1)&&  \nonumber \label{eq:HQE}
\end{eqnarray}
The short distance quantity $\eta_A$ relates HQET  to QCD and is known
to two loop   order\cite{Czarnecki}.    The $\ell$'s are   the  matrix
elements in HQET which we determine from lattice QCD.

We  construct double  ratios\cite{BtoDstar}   from the lattice  matrix
elements     $\braOket{D}{V_4}{B}$,      $\braOket{D^*}{A_1}{B}$   and
$\braOket{D^*}{V_4}{B^*}$.    The matching  between  lattice  and HQET
currents   is known  to  one  loop  order\cite{Harada}. By varying the
``charm''  and ``bottom'' quark masses  in the lattice simulations, we
are able to extract all of the  required $\ell$'s in Eq.~\ref{eq:HQE},
except $\ell_-^{(3)}$.   The  truncation  uncertainty in our  $\hAone$
determination is the size of this unknown term plus higher order terms
in the heavy quark expansion.

\section{RESULT FOR $\mathbf{\hAone}$}

We find
\begin{equation} \nonumber
\qquad\qquad\TheCombinedResult \;\FullStop
\end{equation}
The first uncertainty is the statistical error  added in quadrature to
the uncertainty from fitting procedures.    The second uncertainty  is
the combined error, in  quadrature, from the  other sources  listed in
Tab.~\ref{tab:budget}.

\ErrBudget

\subsection{Systematic Uncertainties}
\BulletItem{Statistics and fitting procedures}.
The statistical error   in   our  result   is after   the   ``chiral''
extrapolation  of  the spectator quark.   The  fitting procedure error
includes  the  effect of excited state  contamination  and also bounds
variations in $\hAone$ from alternate plausible time ranges in fitting
three-point ratio plateaus.  Minimum Chi-square fits were obtained for
all  fits.   Fits  include the data   correlation  matrix  and produce
bootstrap error  determinations.   Small poorly determined eigenvalues
of correlation matrices were rejected in a SVD decomposition.

\BulletItem{Adjustment of $\mathbf{m_c}$ and $\mathbf{m_b}$}.
The charm and bottom masses are determined by adjusting the bare quark
mass  until the kinetic mass of  a lattice meson  matches the physical
$D_s$  and $B_s$ masses  respectively.  These kinetic masses tended to
be quite noisy.   Our uncertainties are  taken to  encompass our $m_c$
and $m_b$    determinations  from  quarkonia. The   agreement  between
heavy-light  and -onia quark mass determinations  is  better for charm
than bottom.  Hence, our charm quark mass has the smaller uncertainty.

\BulletItem{Matching, $\mathbf{\alpha_s^2}$}.
Although the matching  between  HQET and QCD  is   known to two   loop
order\cite{Czarnecki}, the matching   between lattice and QCD  is only
known to  one  loop order\cite{Harada}.  Hence,  we match  among HQET,
lattice and  QCD   schemes consistently  to one loop   order, choosing
scales   for       the   QCD   coupling   according     to    the  BLM
prescription\cite{LPT}.      The  uncertainty in Tab.~\ref{tab:budget}
reflects our estimate of non-BLM terms at and beyond two loop order in
the perturbative expansions.

\BulletItem{Undetermined $\mathbf{(\bar{\Lambda}/(2m_Q))^3}$ terms}.
We  did  not  determine   $\ell_-^{(3)}$ in Eq.~\ref{eq:HQE}.   Taking
nominal values for $m_c$ and $m_b$ and $\ell_-^{(3)} \approx
\bar{\Lambda}  \approx 0.5\,\GeV$ we estimate  the size of the unknown
term to be $\sim 0.0017$. This estimate is consistent with the size of
all other $1/m_Q^3$ terms that \emph{are} included in our result.

\BulletItem{Action and currents $\mathbf{\alpha_s(\bar{\Lambda}/(2m_Q))^2}$}.
The action and currents are  tuned to tadpole-improved  tree level.  A
careful  analysis\cite{BtoDstar}  of  the  form  factors   shows  that
remaining corrections are of order $\alpha_s(\bar{\Lambda}/(2m_Q))^2$.
Using       nominal      values,            we                estimate
$\alpha_s(\bar{\Lambda}/(2m_c))^2\approx0.008$.  We also estimate this
uncertainty  by repeating our analysis  using tree  level quark masses
instead  of the ``quasi    one-loop''   masses used in  our   standard
analysis. We estimate the uncertainty in $\hAone$ to be $\pm0.0114$ by
this method.

\FigSpacDep

\BulletItem{Lattice spacing dependence}.
We  computed $\hAone$ for  a  strange-mass  spectator quark at   three
lattice spacings corresponding to  $\beta=6.1$, $5.9$ and $5.7$.   The
lattice spacing dependence  is shown in Figure~\ref{fig:LatSpac}.   We
take the   weighted average (horizontal line    and error envelope) of
values at our two finest lattice spacings as our best determination of
$\hAone$ .  The  value from our coarsest  lattice is not  used in this
average since  it from   suffers  larger  heavy  quark  discretization
errors.  We  do   include  the coarsest   lattice  in   our  bound  on
discretization errors.  This bound is  shown by the  linear fit to all
three points.

\BulletItem{Chiral extrapolation}.
We have studied the spectator mass  dependence for $\hAone$ at our two
finer lattice  spacings.  Figure~\ref{fig:SpectMass} shows  the linear
extrapolation to  the  physical down  quark mass for  $\beta=5.9$.  We
observe a similar slope, in dimensionful units,  for $\beta=6.1$.  Our
best value of $\hAone$ with a  strange-mass spectator quark is shifted
downwards by  the amount shown in the  figure to yield our final value
of $\hAone$ with a   bottom spectator quark.  Statistical  errors  are
increased as indicated in the figure.

\FigPiDep

Randall and Wise\cite{Randall}   have computed pion loop effects  upon
$\hAone$ in Chiral Perturbation Theory.  The curve with the cusp shown
in Fig.~\ref{fig:SpectMass} is this prediction  for the spectator mass
dependence\cite{BtoDstar}.  The departure  from  linear behaviour near
the  down  quark mass  adds an   additional uncertainty  to the chiral
extrapolation.

\BulletItem{Quenched QCD}.
Our result  is obtained in quenched  QCD.  The quenched  coupling runs
incorrectly: the short-distance quantity $\eta_A$ changes by $+0.0050$
when quenched. Long  distance form factors, such  as $h_{A_1}/\eta_A$,
are  also affected by  quenching.  With our method, however, quenching
errors  only affect the deviations from  unity.   Guided by studies of
decay constants $f_B$ and $f_D$, we expect this  error to be less than
$10\%$. Our quenching uncertainty reflects  estimates of both long and
short distance effects.

\section{$\mathbf{\Vcb}$}

Using our result, we find:
\begin{equation}
\renewcommand{\arraystretch}{1.3}
	10^3|V_{cb}| = \left\{
		\begin{array}{ll}
			45.9 \pm 2.4^{+1.8}_{-1.4} & \textrm{CLEO}\,\cite{CLEO} \\
			38.7 \pm 1.8^{+1.5}_{-1.2} & \textrm{LEP}\,\cite{LEP-Vcb-WG} \\
			39.3 \pm 2.5^{+1.6}_{-1.2} & \textrm{Belle}\,\cite{Belle} \\
		\end{array} \right. ,
\end{equation}
where the  second error results from  adding all  our uncertainties in
quadrature.   This  result   includes a  $+0.007$   QED correction  to
$\hAone$.

\end{document}